\documentclass[fleqn,10pt]{wlscirep}
\usepackage[utf8]{inputenc}
\usepackage[T1]{fontenc}
\usepackage{subcaption}
\usepackage{graphicx}
\usepackage{rotating}
\usepackage{soul}
\title{Detecting ulcerative colitis from colon samples using efficient feature selection and machine learning}

\author[1]{ Hanieh Marvi Khorasani}
\author[2,*]{ Hamid Usefi}
\author[1,*]{ Lourdes Pe\~na-Castillo}
\affil[1]{Department of Computer Science, Memorial University, St. John's, NL, A1B3X5, Canada}
\affil[2]{Department of Mathematics and Statistics, Memorial University, St. John's, NL, A1C5S7, Canada}

\affil[*]{usefi@mun.ca, lourdes@mun.ca}

\begin{abstract}
	
Ulcerative colitis (UC) is one of the most common forms of inflammatory bowel disease (IBD) characterized by inflammation of the mucosal layer of the colon. Diagnosis of UC is based on clinical symptoms, and then confirmed based on endoscopic, histologic and laboratory findings. Feature selection and machine learning have been previously used for creating models to facilitate the diagnosis of certain diseases.  In this work, we used a recently developed feature selection algorithm (DRPT) combined with a support vector machine (SVM) classifier to generate a model to discriminate between healthy subjects and subjects with UC based on the expression values of 32 genes in colon samples. We validated our model with an independent gene expression dataset of colonic samples from subjects in active and inactive periods of UC. Our model perfectly detected all active cases and  had an average precision of 0.62 in the inactive cases. Compared with results reported in previous studies and a model generated by a recently published software for biomarker discovery using machine learning (BioDiscML), our final model for detecting UC shows better performance in terms of average precision.

\end{abstract}
\begin{document}

\flushbottom
\maketitle
\thispagestyle{empty}

\section*{Introduction}

%The Introduction section, of referenced text\cite{Figueredo:2009dg} expands on the background of the work (some overlap with the Abstract is acceptable). The introduction should not include subheadings.

Inflammatory bowel disease (IBD) is a chronic inflammatory condition of the gut with an increasing health burden~\cite{Kaplan:2015kq}. Ulcerative colitis (UC) and Crohn's disease are the two most common forms of chronic IBD with UC being more widespread than Crohn's disease~\cite{Ordas:2012qf}. There is no cure for UC~\cite{Eisenstein:2018kq} and people with the disease alternate between periods of remission (inactive) and active inflammation~\cite{Ordas:2012qf}.  The underlying causes of UC are not completely understood yet, but it is thought to be a combination of genetic, environmental and psychological factors that disrupt the microbial ecosystem of the colon~\cite{Eisenstein:2018kq, Khan:2019qf}. Genome-wide association studies (GWAS)  have identified 240 risk loci for IBD~\cite{Lange:2017eu} and 47 risk loci specifically associated with UC~\cite{Anderson:2011qe}. However, the lower concordance rate in identical twins of 15\% in UC compared with 30\% in Crohn's disease indicates that genetic contribution in UC is weaker than in Crohn's disease~\cite{Conrad:2014yq}. Thus, using gene expression data for  disease diagnostic might be more appropriate for UC than using GWAS data, as it has been done for Crohn's disease~\cite{Romagnoni:2019rt}.

There are several features used for clinical diagnosis of UC including patient symptoms, and laboratory, endoscopic and histological findings~\cite{Conrad:2014yq}.  Boland et at~\cite{Boland:2014ve} carried out a proof-of-concept study for using gene expression measurements from colon samples as a tool for clinical decision support in the treatment of UC. The purpose of Boland et al's study was to discriminate between active and inactive UC cases; even though,  they  only considered gene expression of eight inflammatory genes instead of assessing the discriminatory power of many groups of genes, they concluded that mRNA analysis in UC is a feasible approach to measure quantitative response to therapy. 

Machine learning-based models have a lot of potential to be incorporated into clinical practice~\cite{Shah:2019fp}; specially in the area of medical image analysis~\cite{Esteva:2017qr, McKinney:2020hl}. Supervised machine learning has already proved to be useful in disease diagnosis and prognosis as well as personalized medicine~\cite{molla2004using, xu2019translating}. In IBD, machine learning has been used to classify IBD paediatric patients using endoscopic and histological data~\cite{Mossotto:2017wd},  to distinguish UC colonic samples from control and Crohn's disease colonic samples~\cite{Olsen:2009rz}, and to discriminate between healthy subjects, UC patients, and Crohn's disease patients~\cite{Yuan:2017lq}. Here we apply a machine learning classifier on gene expression data  to generate a model to differentiate UC cases from controls. Unlike previous studies~\cite{Olsen:2009rz,Yuan:2017lq}, we combined a number of independent gene expression data sets instead of using a single data set to train our model, and, by using feature selection, we were able to identify 32 genes out of  thousands genes for which expression measurements were available.  The  expression values of these 32 genes is sufficient to generate a SVM model to effectively discriminate between UC cases and controls. Our proposed model perfectly detected all active cases and had an average precision of 0.62 in the inactive cases.

%************************************************************************

\section*{Methods}

\subsection*{Data gathering}
We searched the NCBI Gene Expression Omnibus database (GEO) for expression profiling studies using colonic samples from UC subjects (in active and inactive state) and controls (healthy donors). We identified five datasets (accession numbers GSE1152~\cite{GSE1152paper}, GSE11223~\cite{GSE11223paper}, GSE22619~\cite{GSE22619paper}, GSE75214~\cite{GSE75214} and GSE9452~\cite{Olsen:2009rz}). As healthy and Crohn's disease subjects were used as controls in GSE9452~\cite{Olsen:2009rz}, this data set was excluded from our study. We used three of the datasets for model selection using 5-fold cross-validation, and left one dataset for independent validation (Table~\ref{datasets_summary}). We partitioned the validation dataset into two datasets: Active UC vs controls, and inactive UC vs controls.

\begin{table}[ht]
		
		\centering
		%	\captionsetup{font=small}
		%\begin{adjustwidth}{-2cm}{}
		\resizebox{\columnwidth}{!}{%
			\begin{tabular}{|c | c | c| c | c | c  | c | c|}
				\hline
				Accession &  $\#$  of   &   $\#$ of   & Description of& Platform &$\#$  of genes  & Usage \\
				
				number &  controls  &   UC cases  &  samples & & (features)  &\\
				
\hline
				&&& && &\\
				GSE1152 \cite{GSE1152paper}  &4 & 4 &  Mucosal biopsies from   & Affymetrix Human Genome  &  19,353 & Model\\
				&&&uninflammed colonic tissues  & U133A Array and Affymetrix  & &selection\\ 
				&&&&Human Genome U133B Array& &\\ 
				&&& && &\\

				GSE11223 \cite{GSE11223paper} & 24 & 25  & Biopsies from    & Agilent-012391 Whole & 18,626 & Model\\
				&&&uninflammed sigmoid colon    &  Human Genome Oligo    & & selection\\ 
				&&&  &Microarray G4112A& &\\ 
				&&& && &\\
				
				GSE22619 \cite{GSE22619paper}, \cite{GSE226192} &10 & 10 & Mucosal  colonic tissue    &  Affymetrix Human Genome&22,189 & Model \\
				&&&  from  discordant twins & U133 Plus 2.0 Array &  & selection\\
				&&& && &\\
				
				GSE75214-active \cite{GSE75214} &11 & 74& Mucosal colonic biopsies from   & Affymetrix Human Gene  & 20,358  & Model \\
				&&&  active UC patients and &1.0 ST Array& & evaluation\\
				&&&  from controls  && &\\
				&&& && &\\
					
				GSE75214-inactive \cite{GSE75214}  &11 & 23 & Mucosal colonic biopsies from & Affymetrix Human Gene & 20,358  & Model \\
				&&& inactive UC patients &1.0 ST Array&& evaluation\\
				&&&  from controls  && &\\
				%	Merged dataset & 77 & 16,313&39 & 38 \\
				
				\hline
				
		\end{tabular}}
		
		%	\end{adjustwidth}
		\caption{\label{datasets_summary}Summary of datasets used in this study.}
	\end{table}

For each dataset, GEO2R~\cite{GEO2R} was used to retrieve the mapping between probe IDs and gene symbols. Probe IDs without a gene mapping were removed from further processing. Expression values for the mapped probe IDs were obtained using the Python package GEOparse~\cite{GEOparse}. The expression values obtained were as provided by the corresponding authors. 

\subsection*{Data Pre-processing}
We performed the following steps for  data pre-processing: (i) Calculating expression values per gene by taking the average of expression values of all probes mapped to the same gene. (i) Handling missing values with K-Nearest Neighbours (KNN) imputation method (KNNImputer) from the ``missingpy'' library in Python~\cite{Troyanskaya:2001yq}. KNNImputer uses KNN to fill in missing values  by utilizing the values from nearest neighbours. We set the number of neighbours to 2 (n-neighbours=2) and we used uniform weight. 

To get our final training datasets we merged datasets  GSE1152, GSE11223, and GSE22619 by taking the genes present in all of them. The merged dataset has 39 UC samples and 38 controls, and 16,313 genes. These same genes were selected from  GSE75214 for validation.  As the range of expression values across all datasets were different, we normalized the expression values of the final merged dataset and validation dataset by calculating Z-scores per sample.

\subsection*{Model Generation}
To create a model to discriminate between UC patients from healthy subjects, we selected the features (genes) using the dimension reduction through perturbation theory (DRPT) feature selection method~\cite{MH}.  Let  $D=[A\mid \textbf{b}]$ be a dataset where $\textbf{b}$ is the class label and $A$ is an $m\times n$ matrix with $n$ columns (genes) and $m$ rows (samples).  There is only a limited number of genes that are associated with the disease, and as such, a majority of genes are considered irrelevant. DRPT considers the solution $\textbf{x}$ of the  linear  system $A\textbf{x}=\textbf{b}$ with the smallest 2-norm.  Hence, $\textbf{b}$ is a sum of $x_i\mathbf{F}_i$ where $\mathbf{F}_i$ is the $i$-th column of $A$. Then each component $x_i$  of $\textbf{x}$ is viewed as an assigned weight to the feature  $\mathbf{F}_i$. So the bigger the $|x_i|$ the more important  $\mathbf{F}_i$ is in connection with  $\textbf{b}$. DRPT then filters out  features whose weights are very small compared to the average of local maximums over  $|x_i|$'s.  After removing irrelevant features, DRPT uses perturbation theory to detect correlations between genes of the reduced dataset. Finally, the remaining genes are sorted based on their entropy.

    Selected features were assessed using 5-fold cross-validation and support vector machines (SVMs) as the classifier. First, we performed DRPT 100 times on the training dataset to generate 100 subsets of features. Then, to find the best subsets, we performed 3 repetitions of stratified 5-fold cross-validation (CV) on the training dataset. We utilized average precision  (AP) as calculated by the function average\_precision\_score from the Python library scikit-learn~\cite{scikit-learn} (version 0.22.1) as the evaluation metric to determine the best subset of genes among those 100 generated subsets. The four subsets with the highest mean AP over the cross-validation folds were chosen for creating the candidate models. For each of four selected subset of features, we created a candidate SVM model using all samples in the training dataset. To generate the models, we used the SVM implementation available in the function SVC with parameter kernel=’linear’ from the Python library scikit-learn. To evaluate the prediction performance of each of the ten models, we validated it on the GSE75214-active and GSE75214-inactive datasets. In this step, we utilized the precision-recall curve (PRC) to assess the performance of the candidate models on unseen data. An additional candidate model was created using the  most frequently selected genes.

\subsection*{BioDiscML}
	BioDiscML~\cite{BioDiscML} is a biomarker discovery software that uses machine learning methods to analyze biological datasets. To compare the prediction performance of our models with BioDiscML, we ran the software on our training dataset. 2/3 of the samples (N=52) were utilized for training and the remaining 1/3 (N=25) for testing. Since the software generates thousands of models, and we required only one model, we specified the number of best models as 1 in the config file (numberOfBestModels=1). One best model out of all models was created based on the 10-fold cross-validated Area Under Precision-Recall Curve (numberOfBestModelsSortingMetric= TRAIN-10CV-AUPRC) on the train set.  We used Weka 3.8 \cite{weka, wekaUpdate, wekaa} to evaluate the performance of the model generated by BioDiscML, on the GSE75214-active and GSE75214-inactive datasets. Selected features by BioDiscML are C3orf36, ADAM30, SLS6A3, FEZF2, and GCNT3. In order to be able to use the model in Weka, we loaded the training dataset as it was created by BioDiscML, which was one of the outputs of the software. This dataset has six features, including selected genes and class labels, and 52 samples. We also modified our validation datasets by extracting BioDiscML selected features. After loading the training and test dataset in Weka explorer, we loaded the model, and we entered the classifier configuration as ``weka.classifiers.misc.InputMappedClassifier -I -trim -W weka.classifiers.trees.RandomTree -- -K 3 -M 1.0 -V 0.001 -S 1'' which is the classifier's set up in the generated model by BioDiscML. 

%************************************************************************

\section*{Results}

\subsection*{Feature selection reduced significantly the number of genes required to construct a classification model}

We performed DRPT 100 times on the training dataset to select 100 subsets of features. Then we performed 5-fold cross-validation to find the subsets with the highest mean average precision  (AP) over the folds. The range of AP for the 100 subsets is between 0.82 and 0.97, with an average of $0.91 \pm 0.03$. Table \ref{Best_subsets_training} shows the ten subsets with the highest cross-validated AP and the number of selected features (genes) on each subset. On average, DRPT selected $37.55  \pm 8.84$ genes per subset. 
\begin{table}[ht]

%% TO DO - ADD STANDARD DEVIATION TO THIS TABLE
	\centering
	%	\captionsetup{font=small}
	%\begin{adjustwidth}{-2cm}{}
	\begin{tabular}{c c c }
		\hline
		\textbf { Subset} &   \textbf{AP } & \textbf{$\#$ of Features} \\ \hline 
		Subset 10 & 0.97 & 42 \\
		Subset 51 &  0.97 &47 \\
		Subset 58 & 0.97 & 32 \\
		Subset 83 &0 .97  &39 \\
		Subset 5 & 0.96 &  37\\
		Subset 16 & 0.96 &  30\\
		Subset 33 & 0.96 &  27\\
		Subset 55 & 0.96 & 22 \\
		Subset 62 & 0.96 &  46\\
		Subset 74 & 0.96 &50 \\
		
		\hline
	\end{tabular}
	%	\end{adjustwidth}
	\caption{\label{Best_subsets_training} Ten top subsets of genes with the highest cross-validated average AP.}
\end{table}

\subsection*{Top five models are able to perfectly discriminate between active UC patients and controls}

\begin{figure}[tbh]
	\centering
	\includegraphics[width=110mm,scale=0.90]{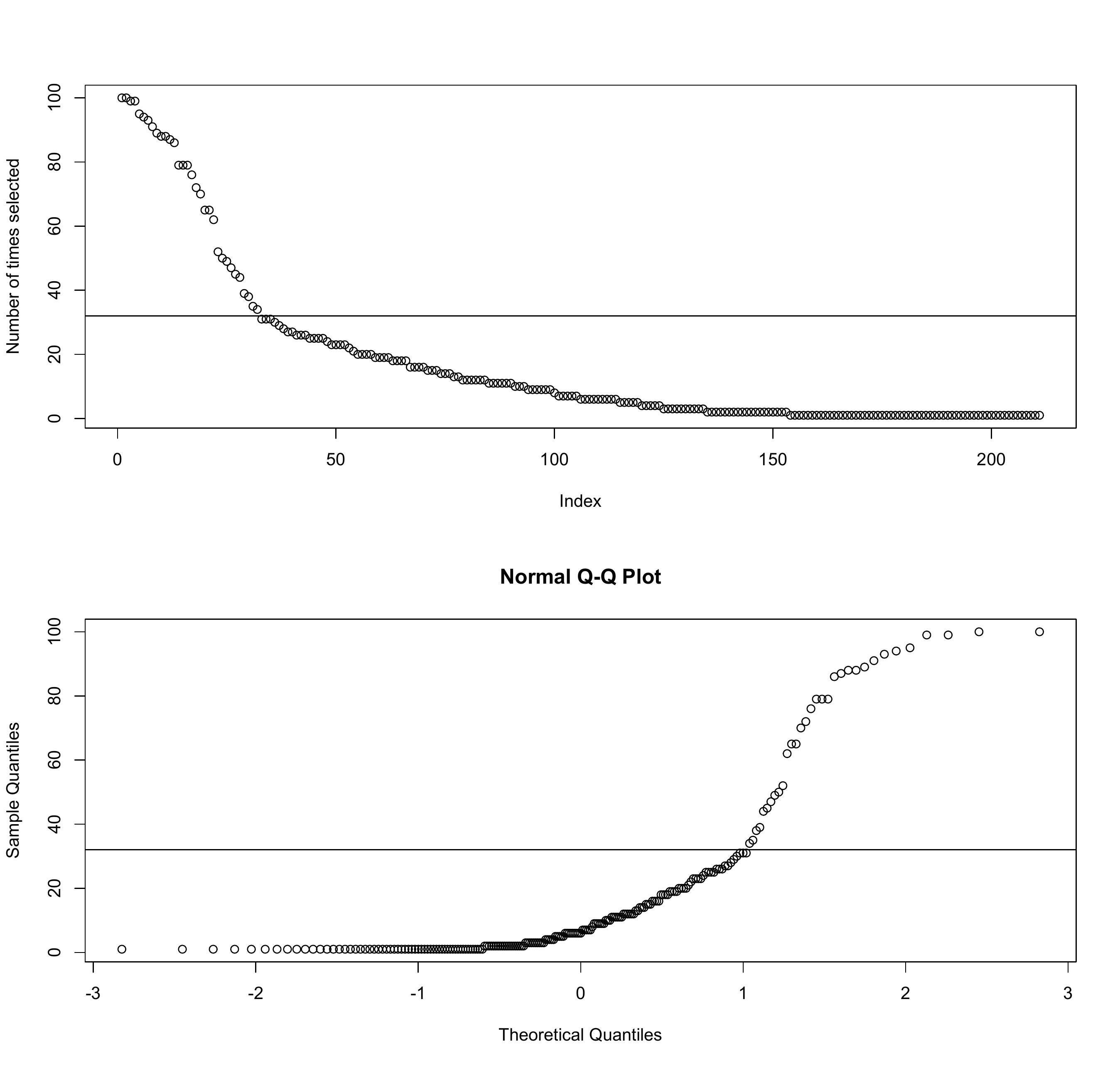}
	\caption{Identifying the most frequently selected genes. Top: Number of times each gene was selected. Genes were  sorted based on the number of times they were selected by DRPT. Bottom: Normal QQ-plot. Horizontal line at 31 indicates the threshold selected to deem a gene as frequently chosen.}
	\label{Q_Q_plot}
\end{figure}

We selected the four top subsets with the highest mean AP, which are subsets 10, 51, 58, and 83 (Table \ref{Best_subsets_training}), and created candidate models based on them. Each candidate model was created using all samples on the training dataset and the features of the corresponding subset. To identify the genes most relevant to discriminate between healthy and UC subjects, we looked at the number of times each gene was selected by DRPT. On 100 DRPT runs, 211 genes were selected at least once. The upper plot on Fig. \ref{Q_Q_plot} shows the number of times each gene was selected, and the lower plot shows the normal quantile-quantile (QQ) plot. Based on this plot, we saw that the observed distribution of the  number of times a gene was selected deviates the most from a Gaussian distribution above 31 times.  We considered the genes selected by DRPT more than 31 times as highly relevant and created a fifth model using 32 genes selected by DRPT at least 32 times over  100 runs.

\begin{figure}[tbh!]
	\centering
	\includegraphics[width=90mm,scale=0.75]{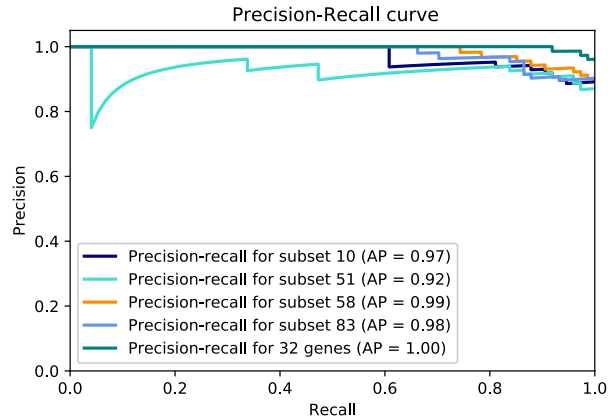}
	\caption{Precision-Recall Curve of Top Selected Subsets on GSE75214-active.}
	\label{Pre-Rec-active}
\end{figure}

\begin{figure}[tbh!]
	\centering
	\includegraphics[width=90mm,scale=0.75]{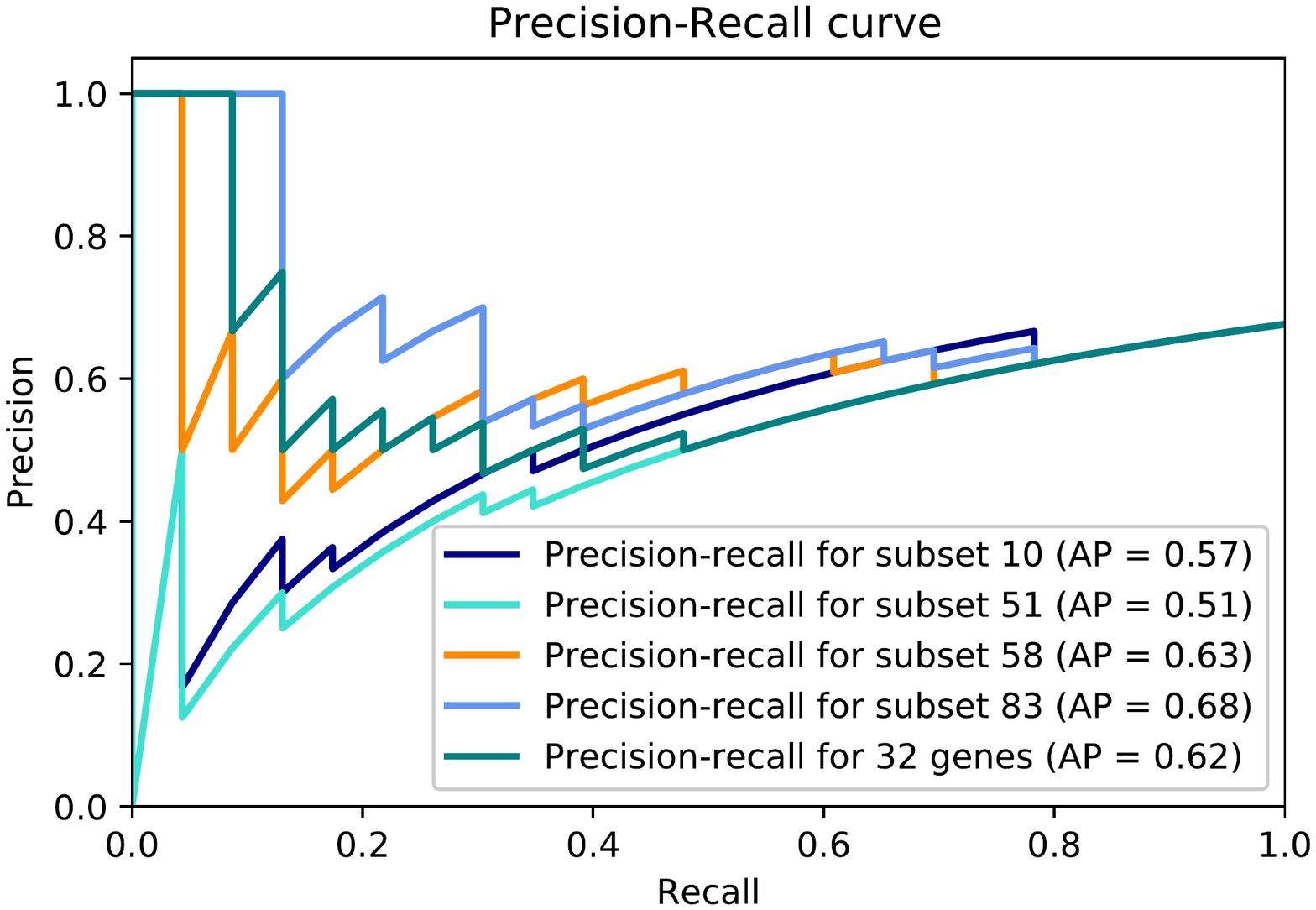}
	\caption{Precision-Recall Curve of Top Selected Subsets on GSE75214-inactive.}
	\label{Pre-Rec-inactive}
\end{figure}

In order to evaluate the prediction performance of the candidate models, each model was tested on the validation datasets, and  PRC was plotted for model assessment (Figs. \ref{Pre-Rec-active}, and \ref{Pre-Rec-inactive}). As the AP approximates the AUPRC~\cite{IntroToML}, we used AP to summarize and compare the performance of these five models. All five candidate models achieved high predictive performance on the validation dataset GSE75214-active with an average AP of $0.97 \pm 0.03$, while the average AP of these five models on the validation dataset GSE75214-inactive was $0.60 \pm 0.06$. The models with the best performance were the model created with the 32 most frequently selected genes and subset 83 with an AP of 1 and 0.68 on GSE75214-active and GSE75214-inactive, respectively. However, based on a Friedman test \cite{Friedman} ($p-value=0.17$), all  five models have comparable performance on the validation datasets. We chose the model generated with the 32 most frequently selected genes as our final model.

\subsection*{Our top models outperformed the model generated by BioDiscML.}

 	 The average AUPRC achieved by the model created by BioDiscML  on both GSE75214-active and GSE75214-inactive datasets was 0.798 and 0.544, respectively. Comparing the performance of our candidate models and the model created by BioDiscML on the two validation datasets, we observed that we achieved better AUPRC on both datasets (AUPRC = 1 on the active dataset, AUPRC = 0.68 on the inactive dataset). 
%However, based on a Friedman test ($p-value=0.12$), there is no statistical significant difference among our five models and the model created by BioDiscML. 
In terms of running time, subset selection by DRPT and final model creation and validation, took 3 minutes, while the running time of BioDiscML to create all the models and output the best final model was 1,890 minutes.

\subsection*{Links between the most frequently selected genes and UC.}

We used  Ensembl REST API (Version 11.0) \cite{ensembl} to find the associated phenotypes with each gene belonging to the subset of the 32 most frequently selected genes (Table \ref{32genes}).  Among these 32 genes, FAM118A is the only one with a known phenotypic association with IBD and its subtypes. The evidence supporting the association of some of the other 31 genes with UC based on phenotype is more indirect. For example,  long term  IBD patients are more susceptible to develop colorectal cancer \cite{Colorectal}, and one of the 32 genes, TFRC, is associated with colorectal cancer. IBD patients are more prone to develop cardio vascular disease which is associated with blood pressure and cholesterol \cite{Cholestrol}, and four of the most frequently selected genes (LIPF, MMP2, DMTN and PPP1CB) are associated with blood pressure and cholesterol.

%%%%%%%%%%%%%%%%%%%%%%%Analyzing genes / Part 1
\begin{sidewaystable}[tbp]
	\begin{center}
	\resizebox{\textwidth}{!}{%
		\begin{tabular}{c l c }
			\hline
			Gene Symbol   & Associated Phenotypes & $\#$ of times \\  
			&& selected\\			\hline 
			
			CWF19L1 & Spinocerebellar ataxia, autosomal recessive 17; depressive disorder, Major & 100 \\ 
			
			FCER2 & Blood protein levels; post bronchodilator FEV1 &100\\
			
			MMP2 & Multicentric Osteolysis-Nodulosis-Arthropathy (MONA) spectrum disorders; cholesterol, HDL; lip and oral cavity carcinoma;  \\& body height; winchester syndrome  & 99 \\
			
			PPP1CB & Noonan Syndrome-like disorder with loose anagen hair 2; Heel bone mineral density; Blood pressure; \\
			& basophils asopathy with developmental delay; short stature and sparse slow-growing hair& 99\\		
				
			RPL23AP32 & Attention deficit disorder with hyperactivity; body Height & 95 \\
	
			ZNF624 & None &94 \\
			
			REG1B &Contrast sensitivity; Body Mass Index & 93 \\
			
			TFRC & Breast ductal adenocarcinoma; esophageal adenocarcinoma; thyroid carcinoma; clear cell renal carcinoma; prostate carcinoma; pancreatic cancer;\\ 
			&   gastric adenocarcinoma; hepatocellular carcinoma;  lung adenocarcinoma; rectal adenocarcinoma; basal cell carcinoma; {\bf colorectal adenocarcinoma};\\
			&  squamous cell lung carcinoma; head and neck squamous cell carcinoma; {\bf colon adenocarcinoma}; iron status biomarkers (transferrin levels);  \\ 
			&  mean corpuscular hemoglobin concentration; red cell distribution width; combined immunodeficiency; red blood cell traits; high light scatter   \\
			& reticulocyte percentage of red cells; reticulocyte fraction of red cells; Immunodeficiency 46 &91 \\

			FAM118A & {\bf Chronic inflammatory diseases} (ankylosing spondylitis, Crohn's disease, psoriasis, primary sclerosing cholangitis, ulcerative colitis); Glucose;  \\ &  Peanut allergy (maternal genetic effects); Heel bone mineral density & 89\\
			
			CFHR2 & Macular degeneration; blood protein levels; feeling miserable; alanine aminotransferase (ALT) levels after remission induction therapy    \\& in acute lymphoblastic leukaemia (ALL); asthma  &88 \\
		
			KRT8 & Cirrhosis; familial cirrhosis; hepatitis C virus; susceptibility to, cirrhosis, cryptogenic cirrhosis, noncryptogenic cirrhosis;  \\ &  susceptibility to, gamma glutamyl transferase levels, cancer (pleiotropy) &88\\
			 
			PRELID1 & Body fat distribution; heel bone mineral density; activated partial thromboplastin time &87 \\ 
		
			ZNF92 & None & 86\\
			
			ABHD2 & Itch intensity from mosquito bite adjusted by bite size; gut microbiota; Obesity-related traits; coronary artery disease; advanced    \\& age related macular degeneration; squamous cell lung carcinoma; pulse pressure &79 \\
			
			C16orf89 & None & 79\\
			
			CAB39L & Hemoglobin S; erythrocyte count; pancreatic neoplasms & 79  \\
			
			SPATC1L & None & 76\\
			
			DUOXA2 & Familial thyroid dyshormonogenesis; thyroglobulin synthesis defect &72   \\
			
			MESP1 & None & 70 \\
			
			MAML3 & Social science traits; intelligence (MTAG); chronic mucus hypersecretion; borderline personality disorder; congenital heart malformation & 65 \\
			
			PITX2 & Axenfeld-Rieger syndrome; ring dermoid of cornea; iridogoniodygenesis type 2; peters anomaly; familial atrial fibrillation; rieger anomaly; stroke; \\&  ischemic stroke;  cataract;  PITX2-related eye abnormalities; phosphorus; cognitive decline rate in late mild cognitive impairment;   \\& creatinine; intraocular pressure; incident atrial fibrillation; wolff-parkinson-white pattern; \\& parkinson disease; early onset atrial fibrillation; anterior segment sygenesis 4 & 65 \\

			DMTN & Total cholesterol levels; LDL cholesterol & 62 \\
			
			ASF1B & None &  52\\
			
			PGF & Mood instability; blood protein levels & 50  \\
			
			BEX4 & None & 49\\

			ODF1 & Body weight; body mass index; glucose; IgA nephropathy; Chronic lymphocytic leukaemia;  type 2 diabetes; erythrocyte indices & 47  \\
			
			PTGR1 &Body height; menarche; monocyte count; blood protein levels &45 \\
			
			ZNF35 & None & 44\\
			
			LIPF & Maximal midexpiratory flow rate; blood protein levels; respiratory function tests; blood pressure & 39\\
			
			SLC25A13 & Citrullinemia type II; neonatal intrahepatic cholestasis due to citrin deficiency; citrin deficiency; citrullinemia type I; bone mineral density & 38   \\
			
			BARX2 & Type 2 diabetes; breast cancer; night sleep phenotypes; response to cyclophosphamide in systemic lupus erythematosus with lupus nephritis; stroke & 35  \\
			
			C2orf42 & None & 34\\

		\end{tabular}}
	\caption{ Phenotypes associated with the 32 most frequently selected genes by DRPT.}
		\label{32genes}
	\end{center}	
	%	\end{adjustwidth}
\end{sidewaystable}

We looked at whether some of the 32 most frequently selected genes contained any of the 241 known IBD-associated SNPs~\cite{Lange:2017eu}. To do this, we utilized Ensembl's BioMart  \cite{Biomart} website (Ensembl Release version 98 - September 2019) to retrieve the genomic location of the 32 genes.  We then used the intersectBed utility in BEDtools~\cite{Quinlan2010} to find any overlap between the 241 IDB risk loci and the genomic location of the 32 genes. None of the IBD-associated SNPs was located on our 32 genes. Similarly, gene set enrichment analysis found no enriched GO term or pathway among these 32 genes.  Additionally, these 32 genes are not listed as top differentially expressed genes in previous studies on UC~\cite{Roman:2013ph,Song:2018it}.

We searched the literature for links between the 32 genes and UC, and we found the following. MMP2 expression has been found significantly increased in colorectal neoplasia in a mouse model of UC~\cite{Schwegmann:2016kq} and  MMP2 levels are elevated in IBD~\cite{Oliveira:2014qf}. TFRC has been found to have an anti-inflammatory effect on a murine colitis model~\cite{Shin:2014vn}. KRT8 genetic variants have been observed in IBD patients and it was suggested that these variants are a risk factor for IBD~\cite{Owens:2004ys}. DUOXA2 has been shown to be critical in the production of hydrogen peroxide within the colon and to be upregulated in active UC~\cite{MacFie:2014rm}.

%%%%%%%%%%%%%%%%%%%%%%%%%%%%%
\clearpage
%The Discussion should be succinct and must not contain subheadings.
\section*{Discussion}

In a previous study where machine learning was employed to perform a risk assessment for Crohn's disease and UC using GWAS data~\cite{Large},  a two-step feature selection strategy was used on a dataset containing 17,000 Crohn's disease cases, 13,000 UC cases, and 22,000 controls with 178,822 SNPs. In that study, Wei et al reduced the number of features by filtering out SNPs with $p$-values greater than $10^{-4}$ and then applied a penalized feature selection with $L_1$ penalty to select a subset of SNPs. We decided against filtering out genes based on an arbitrary $p$-value of statistical significance of differential expression, as  researchers are strongly advised against the use of $p$-values and statistical significance in relation to the null-hypothesis \cite{amrhein2019scientists, wasserstein2019moving}. To avoid systematic experimental bias on the training data, we used three transcriptomic datasets from three separated studies, and used an independent dataset to validate our top performing models. 

Our 32-gene model achieved AP of 1 and 0.62 discriminating active UC patients from healthy donors, and inactive UC patients from healthy donors, respectively. We found direct or indirect links to UC for about a quarter of the 32 most frequently chosen genes. The remaining genes should be further investigated to find associations with UC.  To put the performance of our 32-gene model into perspective, we looked at previous studies applying machine learning to create models for the diagnostic of UC.   Maeda et al~\cite{Maeda:2019db} extracted 312 features from endocystoscopy images to train a SVM to classify UC patients as active or healing. This approach achieve 90\% precision at 74\% recall; which is lower than the one achieved by our 32-gene model (Figs.~\ref{Pre-Rec-active}, and \ref{Pre-Rec-inactive}). Yuan et al~\cite{Yuan:2017lq} applied incremental feature selection and a SMO classifier (a type of SVM) on gene expression data from blood samples to discriminate between healthy subjects, UC patients, and Crohn's disease patients. The 10-fold cross-validation accuracy of their best model using the expression values of 1170 genes to classify UC patients was 92.31\%, while our method obtained better accuracy than this with substantially less number of genes. In terms of potential for clinical translation of a machine learning-based model, a model requiring to quantify the gene expression levels of fewer genes is more suitable for the development of a new diagnostic test than one requiring the quantification of the expression levels of thousands of genes. 
 Using an efficient feature selection method such as DRPT and a SVM-classifier on gene expression data, we generated a model that could facilitate the diagnosis of UC from expression measurements of 32 genes from colonic samples.

%\bibliography{sample.bib}

%\bibliography{mainSR}

\begin{thebibliography}{10}
	\urlstyle{rm}
	\expandafter\ifx\csname url\endcsname\relax
	\def\url#1{\texttt{#1}}\fi
	\expandafter\ifx\csname urlprefix\endcsname\relax\def\urlprefix{URL }\fi
	\expandafter\ifx\csname doiprefix\endcsname\relax\def\doiprefix{DOI: }\fi
	\providecommand{\bibinfo}[2]{#2}
	\providecommand{\eprint}[2][]{\url{#2}}
	
	\bibitem{Kaplan:2015kq}
	\bibinfo{author}{Kaplan, G.~G.}
	\newblock \bibinfo{journal}{\bibinfo{title}{The global burden of {IBD}: from
			2015 to 2025}}.
	\newblock {\emph{\JournalTitle{Nat Rev Gastroenterol Hepatol}}}
	\textbf{\bibinfo{volume}{12}}, \bibinfo{pages}{720--7},
	\doiprefix\url{10.1038/nrgastro.2015.150} (\bibinfo{year}{2015}).
	
	\bibitem{Ordas:2012qf}
	\bibinfo{author}{Ord{\'a}s, I.}, \bibinfo{author}{Eckmann, L.},
	\bibinfo{author}{Talamini, M.}, \bibinfo{author}{Baumgart, D.~C.} \&
	\bibinfo{author}{Sandborn, W.~J.}
	\newblock \bibinfo{journal}{\bibinfo{title}{Ulcerative colitis}}.
	\newblock {\emph{\JournalTitle{Lancet}}} \textbf{\bibinfo{volume}{380}},
	\bibinfo{pages}{1606--19}, \doiprefix\url{10.1016/S0140-6736(12)60150-0}
	(\bibinfo{year}{2012}).
	
	\bibitem{Eisenstein:2018kq}
	\bibinfo{author}{Eisenstein, M.}
	\newblock \bibinfo{journal}{\bibinfo{title}{Ulcerative colitis: towards
			remission}}.
	\newblock {\emph{\JournalTitle{Nature}}} \textbf{\bibinfo{volume}{563}},
	\bibinfo{pages}{S33}, \doiprefix\url{10.1038/d41586-018-07276-2}
	(\bibinfo{year}{2018}).
	
	\bibitem{Khan:2019qf}
	\bibinfo{author}{Khan, I.} \emph{et~al.}
	\newblock \bibinfo{journal}{\bibinfo{title}{Alteration of gut microbiota in
			inflammatory bowel disease ({IBD}): Cause or consequence? {IBD} treatment
			targeting the gut microbiome}}.
	\newblock {\emph{\JournalTitle{Pathogens}}} \textbf{\bibinfo{volume}{8}},
	\doiprefix\url{10.3390/pathogens8030126} (\bibinfo{year}{2019}).
	
	\bibitem{Lange:2017eu}
	\bibinfo{author}{de~Lange, K.~M.} \emph{et~al.}
	\newblock \bibinfo{journal}{\bibinfo{title}{Genome-wide association study
			implicates immune activation of multiple integrin genes in inflammatory bowel
			disease}}.
	\newblock {\emph{\JournalTitle{Nat Genet}}} \textbf{\bibinfo{volume}{49}},
	\bibinfo{pages}{256--261}, \doiprefix\url{10.1038/ng.3760}
	(\bibinfo{year}{2017}).
	
	\bibitem{Anderson:2011qe}
	\bibinfo{author}{Anderson, C.~A.} \emph{et~al.}
	\newblock \bibinfo{journal}{\bibinfo{title}{Meta-analysis identifies 29
			additional ulcerative colitis risk loci, increasing the number of confirmed
			associations to 47}}.
	\newblock {\emph{\JournalTitle{Nat Genet}}} \textbf{\bibinfo{volume}{43}},
	\bibinfo{pages}{246--52}, \doiprefix\url{10.1038/ng.764}
	(\bibinfo{year}{2011}).
	
	\bibitem{Conrad:2014yq}
	\bibinfo{author}{Conrad, K.}, \bibinfo{author}{Roggenbuck, D.} \&
	\bibinfo{author}{Laass, M.~W.}
	\newblock \bibinfo{journal}{\bibinfo{title}{Diagnosis and classification of
			ulcerative colitis}}.
	\newblock {\emph{\JournalTitle{Autoimmun Rev}}} \textbf{\bibinfo{volume}{13}},
	\bibinfo{pages}{463--6}, \doiprefix\url{10.1016/j.autrev.2014.01.028}
	(\bibinfo{year}{2014}).
	
	\bibitem{Romagnoni:2019rt}
	\bibinfo{author}{Romagnoni, A.} \emph{et~al.}
	\newblock \bibinfo{journal}{\bibinfo{title}{Comparative performances of machine
			learning methods for classifying {C}rohn disease patients using genome-wide
			genotyping data}}.
	\newblock {\emph{\JournalTitle{Sci Rep}}} \textbf{\bibinfo{volume}{9}},
	\bibinfo{pages}{10351}, \doiprefix\url{10.1038/s41598-019-46649-z}
	(\bibinfo{year}{2019}).
	
	\bibitem{Boland:2014ve}
	\bibinfo{author}{Boland, B.~S.} \emph{et~al.}
	\newblock \bibinfo{journal}{\bibinfo{title}{Validated gene expression biomarker
			analysis for biopsy-based clinical trials in ulcerative colitis}}.
	\newblock {\emph{\JournalTitle{Aliment Pharmacol Ther}}}
	\textbf{\bibinfo{volume}{40}}, \bibinfo{pages}{477--85},
	\doiprefix\url{10.1111/apt.12862} (\bibinfo{year}{2014}).
	
	\bibitem{Shah:2019fp}
	\bibinfo{author}{Shah, P.} \emph{et~al.}
	\newblock \bibinfo{journal}{\bibinfo{title}{Artificial intelligence and machine
			learning in clinical development: a translational perspective}}.
	\newblock {\emph{\JournalTitle{NPJ Digit Med}}} \textbf{\bibinfo{volume}{2}},
	\bibinfo{pages}{69}, \doiprefix\url{10.1038/s41746-019-0148-3}
	(\bibinfo{year}{2019}).
	
	\bibitem{Esteva:2017qr}
	\bibinfo{author}{Esteva, A.} \emph{et~al.}
	\newblock \bibinfo{journal}{\bibinfo{title}{Dermatologist-level classification
			of skin cancer with deep neural networks}}.
	\newblock {\emph{\JournalTitle{Nature}}} \textbf{\bibinfo{volume}{542}},
	\bibinfo{pages}{115--118}, \doiprefix\url{10.1038/nature21056}
	(\bibinfo{year}{2017}).
	
	\bibitem{McKinney:2020hl}
	\bibinfo{author}{McKinney, S.~M.} \emph{et~al.}
	\newblock \bibinfo{journal}{\bibinfo{title}{International evaluation of an {AI}
			system for breast cancer screening}}.
	\newblock {\emph{\JournalTitle{Nature}}} \textbf{\bibinfo{volume}{577}},
	\bibinfo{pages}{89--94}, \doiprefix\url{10.1038/s41586-019-1799-6}
	(\bibinfo{year}{2020}).
	
	\bibitem{molla2004using}
	\bibinfo{author}{Molla, M.}, \bibinfo{author}{Waddell, M.},
	\bibinfo{author}{Page, D.} \& \bibinfo{author}{Shavlik, J.}
	\newblock \bibinfo{journal}{\bibinfo{title}{Using machine learning to design
			and interpret gene-expression microarrays}}.
	\newblock {\emph{\JournalTitle{AI Magazine}}} \textbf{\bibinfo{volume}{25}},
	\bibinfo{pages}{23--23} (\bibinfo{year}{2004}).
	
	\bibitem{xu2019translating}
	\bibinfo{author}{Xu, J.} \emph{et~al.}
	\newblock \bibinfo{journal}{\bibinfo{title}{Translating cancer genomics into
			precision medicine with artificial intelligence: applications, challenges and
			future perspectives}}.
	\newblock {\emph{\JournalTitle{Human genetics}}}
	\textbf{\bibinfo{volume}{138}}, \bibinfo{pages}{109--124}
	(\bibinfo{year}{2019}).
	
	\bibitem{Mossotto:2017wd}
	\bibinfo{author}{Mossotto, E.} \emph{et~al.}
	\newblock \bibinfo{journal}{\bibinfo{title}{Classification of paediatric
			inflammatory bowel disease using machine learning}}.
	\newblock {\emph{\JournalTitle{Sci Rep}}} \textbf{\bibinfo{volume}{7}},
	\bibinfo{pages}{2427}, \doiprefix\url{10.1038/s41598-017-02606-2}
	(\bibinfo{year}{2017}).
	
	\bibitem{Olsen:2009rz}
	\bibinfo{author}{Olsen, J.} \emph{et~al.}
	\newblock \bibinfo{journal}{\bibinfo{title}{Diagnosis of ulcerative colitis
			before onset of inflammation by multivariate modeling of genome-wide gene
			expression data}}.
	\newblock {\emph{\JournalTitle{Inflamm Bowel Dis}}}
	\textbf{\bibinfo{volume}{15}}, \bibinfo{pages}{1032--8},
	\doiprefix\url{10.1002/ibd.20879} (\bibinfo{year}{2009}).
	
	\bibitem{Yuan:2017lq}
	\bibinfo{author}{Yuan, F.}, \bibinfo{author}{Zhang, Y.-H.},
	\bibinfo{author}{Kong, X.-Y.} \& \bibinfo{author}{Cai, Y.-D.}
	\newblock \bibinfo{journal}{\bibinfo{title}{Identification of candidate genes
			related to inflammatory bowel disease using minimum redundancy maximum
			relevance, incremental feature selection, and the shortest-path approach}}.
	\newblock {\emph{\JournalTitle{Biomed Res Int}}}
	\textbf{\bibinfo{volume}{2017}}, \bibinfo{pages}{5741948},
	\doiprefix\url{10.1155/2017/5741948} (\bibinfo{year}{2017}).
	
	\bibitem{GSE1152paper}
	\bibinfo{author}{Zahn, A.} \emph{et~al.}
	\newblock \bibinfo{journal}{\bibinfo{title}{Aquaporin-8 expression is reduced
			in ileum and induced in colon of patients with ulcerative colitis}}.
	\newblock {\emph{\JournalTitle{World Journal of Gastroenterology: WJG}}}
	\textbf{\bibinfo{volume}{13}}, \bibinfo{pages}{1687} (\bibinfo{year}{2007}).
	
	\bibitem{GSE11223paper}
	\bibinfo{author}{Noble, C.~L.} \emph{et~al.}
	\newblock \bibinfo{journal}{\bibinfo{title}{Regional variation in gene
			expression in the healthy colon is dysregulated in ulcerative colitis}}.
	\newblock {\emph{\JournalTitle{Gut}}} \textbf{\bibinfo{volume}{57}},
	\bibinfo{pages}{1398--1405} (\bibinfo{year}{2008}).
	
	\bibitem{GSE22619paper}
	\bibinfo{author}{Lepage, P.} \emph{et~al.}
	\newblock \bibinfo{journal}{\bibinfo{title}{Twin study indicates loss of
			interaction between microbiota and mucosa of patients with ulcerative
			colitis}}.
	\newblock {\emph{\JournalTitle{Gastroenterology}}}
	\textbf{\bibinfo{volume}{141}}, \bibinfo{pages}{227--236}
	(\bibinfo{year}{2011}).
	
	\bibitem{GSE75214}
	\bibinfo{author}{Vancamelbeke, M.} \emph{et~al.}
	\newblock \bibinfo{journal}{\bibinfo{title}{Genetic and transcriptomic bases of
			intestinal epithelial barrier dysfunction in inflammatory bowel disease}}.
	\newblock {\emph{\JournalTitle{Inflammatory bowel diseases}}}
	\textbf{\bibinfo{volume}{23}}, \bibinfo{pages}{1718--1729}
	(\bibinfo{year}{2017}).
	
	\bibitem{GSE226192}
	\bibinfo{author}{H{\"a}sler, R.} \emph{et~al.}
	\newblock \bibinfo{journal}{\bibinfo{title}{A functional methylome map of
			ulcerative colitis}}.
	\newblock {\emph{\JournalTitle{Genome research}}}
	\textbf{\bibinfo{volume}{22}}, \bibinfo{pages}{2130--2137}
	(\bibinfo{year}{2012}).
	
	\bibitem{GEO2R}
	\bibinfo{author}{Barrett, T.} \emph{et~al.}
	\newblock \bibinfo{journal}{\bibinfo{title}{{NCBI GEO}: archive for functional
			genomics data sets---update}}.
	\newblock {\emph{\JournalTitle{Nucleic acids research}}}
	\textbf{\bibinfo{volume}{41}}, \bibinfo{pages}{D991--D995}
	(\bibinfo{year}{2012}).
	
	\bibitem{GEOparse}
	\bibinfo{author}{Gumienny, R.}
	\newblock \bibinfo{title}{{GEOparse}}.
	\newblock \bibinfo{howpublished}{\url{https://pypi.org/project/GEOparse/}}.
	
	\bibitem{Troyanskaya:2001yq}
	\bibinfo{author}{Troyanskaya, O.} \emph{et~al.}
	\newblock \bibinfo{journal}{\bibinfo{title}{Missing value estimation methods
			for {DNA} microarrays}}.
	\newblock {\emph{\JournalTitle{Bioinformatics}}} \textbf{\bibinfo{volume}{17}},
	\bibinfo{pages}{520--5}, \doiprefix\url{10.1093/bioinformatics/17.6.520}
	(\bibinfo{year}{2001}).
	
	\bibitem{MH}
	\bibinfo{author}{Afshar, M.} \& \bibinfo{author}{Usefi, H.}
	\newblock \bibinfo{title}{High-{D}imensional {F}eature {S}election for
		{G}enomics {D}atasets}.
	\newblock \bibinfo{note}{ArXiv2002.12104}.
	
	\bibitem{scikit-learn}
	\bibinfo{author}{Pedregosa, F.} \emph{et~al.}
	\newblock \bibinfo{journal}{\bibinfo{title}{Scikit-learn: Machine learning in
			{P}ython}}.
	\newblock {\emph{\JournalTitle{Journal of Machine Learning Research}}}
	\textbf{\bibinfo{volume}{12}}, \bibinfo{pages}{2825--2830}
	(\bibinfo{year}{2011}).
	
	\bibitem{BioDiscML}
	\bibinfo{author}{Leclercq, M.} \emph{et~al.}
	\newblock \bibinfo{journal}{\bibinfo{title}{Large-scale automatic feature
			selection for biomarker discovery in high-dimensional omics data}}.
	\newblock {\emph{\JournalTitle{Frontiers in genetics}}}
	\textbf{\bibinfo{volume}{10}}, \bibinfo{pages}{452} (\bibinfo{year}{2019}).
	
	\bibitem{weka}
	\bibinfo{author}{Holmes, G.}, \bibinfo{author}{Donkin, A.} \&
	\bibinfo{author}{Witten, I.~H.}
	\newblock \bibinfo{title}{Weka: A machine learning workbench}.
	\newblock In \emph{\bibinfo{booktitle}{Proceedings of ANZIIS '94 - Australian
			New Zealand Intelligent Information Systems Conference}},
	\bibinfo{pages}{357--361} (\bibinfo{year}{1994}).
	
	\bibitem{wekaUpdate}
	\bibinfo{author}{Hall, M.} \emph{et~al.}
	\newblock \bibinfo{journal}{\bibinfo{title}{The weka data mining software: an
			update}}.
	\newblock {\emph{\JournalTitle{ACM SIGKDD explorations newsletter}}}
	\textbf{\bibinfo{volume}{11}}, \bibinfo{pages}{10--18}
	(\bibinfo{year}{2009}).
	
	\bibitem{wekaa}
	\bibinfo{author}{Witten, I.~H.}, \bibinfo{author}{Frank, E.},
	\bibinfo{author}{Hall, M.~A.} \& \bibinfo{author}{Pal, C.~J.}
	\newblock \emph{\bibinfo{title}{Data Mining: Practical machine learning tools
			and techniques}} (\bibinfo{publisher}{Morgan Kaufmann},
	\bibinfo{year}{2016}).
	
	\bibitem{IntroToML}
	\bibinfo{author}{M{\"u}ller, A.~C.}, \bibinfo{author}{Guido, S.} \emph{et~al.}
	\newblock \emph{\bibinfo{title}{Introduction to machine learning with Python: a
			guide for data scientists}} (\bibinfo{publisher}{" O'Reilly Media, Inc."},
	\bibinfo{year}{2016}).
	
	\bibitem{Friedman}
	\bibinfo{author}{Dem{\v{s}}ar, J.}
	\newblock \bibinfo{journal}{\bibinfo{title}{Statistical comparisons of
			classifiers over multiple data sets}}.
	\newblock {\emph{\JournalTitle{Journal of Machine learning research}}}
	\textbf{\bibinfo{volume}{7}}, \bibinfo{pages}{1--30} (\bibinfo{year}{2006}).
	
	\bibitem{ensembl}
	\bibinfo{author}{Yates, A.} \emph{et~al.}
	\newblock \bibinfo{journal}{\bibinfo{title}{The {E}nsembl {REST API}: Ensembl
			data for any language}}.
	\newblock {\emph{\JournalTitle{Bioinformatics}}} \textbf{\bibinfo{volume}{31}},
	\bibinfo{pages}{143--145} (\bibinfo{year}{2014}).
	
	\bibitem{Colorectal}
	\bibinfo{author}{Kim, E.~R.} \& \bibinfo{author}{Chang, D.~K.}
	\newblock \bibinfo{journal}{\bibinfo{title}{Colorectal cancer in inflammatory
			bowel disease: the risk, pathogenesis, prevention and diagnosis}}.
	\newblock {\emph{\JournalTitle{World journal of gastroenterology: WJG}}}
	\textbf{\bibinfo{volume}{20}}, \bibinfo{pages}{9872} (\bibinfo{year}{2014}).
	
	\bibitem{Cholestrol}
	\bibinfo{author}{Schulte, D.} \emph{et~al.}
	\newblock \bibinfo{journal}{\bibinfo{title}{Small dense {LDL} cholesterol in
			human subjects with different chronic inflammatory diseases}}.
	\newblock {\emph{\JournalTitle{Nutrition, Metabolism and Cardiovascular
				Diseases}}} \textbf{\bibinfo{volume}{28}}, \bibinfo{pages}{1100--1105}
	(\bibinfo{year}{2018}).
	
	\bibitem{Biomart}
	\bibinfo{author}{Smedley, D.} \emph{et~al.}
	\newblock \bibinfo{journal}{\bibinfo{title}{Biomart--biological queries made
			easy}}.
	\newblock {\emph{\JournalTitle{BMC genomics}}} \textbf{\bibinfo{volume}{10}},
	\bibinfo{pages}{22} (\bibinfo{year}{2009}).
	
	\bibitem{Quinlan2010}
	\bibinfo{author}{Quinlan, A.~R.} \& \bibinfo{author}{Hall, I.~M.}
	\newblock \bibinfo{journal}{\bibinfo{title}{{BEDTools}: a flexible suite of
			utilities for comparing genomic features}}.
	\newblock {\emph{\JournalTitle{Bioinformatics}}} \textbf{\bibinfo{volume}{26}},
	\bibinfo{pages}{841--2}, \doiprefix\url{10.1093/bioinformatics/btq033}
	(\bibinfo{year}{2010}).
	
	\bibitem{Roman:2013ph}
	\bibinfo{author}{Rom{\'a}n, J.} \emph{et~al.}
	\newblock \bibinfo{journal}{\bibinfo{title}{Evaluation of responsive gene
			expression as a sensitive and specific biomarker in patients with ulcerative
			colitis}}.
	\newblock {\emph{\JournalTitle{Inflamm Bowel Dis}}}
	\textbf{\bibinfo{volume}{19}}, \bibinfo{pages}{221--9},
	\doiprefix\url{10.1002/ibd.23020} (\bibinfo{year}{2013}).
	
	\bibitem{Song:2018it}
	\bibinfo{author}{Song, R.} \emph{et~al.}
	\newblock \bibinfo{journal}{\bibinfo{title}{Identification and analysis of key
			genes associated with ulcerative colitis based on {DNA} microarray data}}.
	\newblock {\emph{\JournalTitle{Medicine (Baltimore)}}}
	\textbf{\bibinfo{volume}{97}}, \bibinfo{pages}{e10658},
	\doiprefix\url{10.1097/MD.0000000000010658} (\bibinfo{year}{2018}).
	
	\bibitem{Schwegmann:2016kq}
	\bibinfo{author}{Schwegmann, K.} \emph{et~al.}
	\newblock \bibinfo{journal}{\bibinfo{title}{Detection of early murine
			colorectal cancer by {MMP}-2/-9-guided fluorescence endoscopy}}.
	\newblock {\emph{\JournalTitle{Inflamm Bowel Dis}}}
	\textbf{\bibinfo{volume}{22}}, \bibinfo{pages}{82--91},
	\doiprefix\url{10.1097/MIB.0000000000000605} (\bibinfo{year}{2016}).
	
	\bibitem{Oliveira:2014qf}
	\bibinfo{author}{Oliveira, L. G.~d.} \emph{et~al.}
	\newblock \bibinfo{journal}{\bibinfo{title}{Positive correlation between
			disease activity index and matrix metalloproteinases activity in a rat model
			of colitis}}.
	\newblock {\emph{\JournalTitle{Arq Gastroenterol}}}
	\textbf{\bibinfo{volume}{51}}, \bibinfo{pages}{107--12},
	\doiprefix\url{10.1590/s0004-28032014000200007} (\bibinfo{year}{2014}).
	
	\bibitem{Shin:2014vn}
	\bibinfo{author}{Shin, J.-S.} \emph{et~al.}
	\newblock \bibinfo{journal}{\bibinfo{title}{Anti-inflammatory effect of a
			standardized triterpenoid-rich fraction isolated from {R}ubus coreanus on
			dextran sodium sulfate-induced acute colitis in mice and {LPS}-induced
			macrophages}}.
	\newblock {\emph{\JournalTitle{J Ethnopharmacol}}} \textbf{\bibinfo{volume}{158
			Pt A}}, \bibinfo{pages}{291--300}, \doiprefix\url{10.1016/j.jep.2014.10.044}
	(\bibinfo{year}{2014}).
	
	\bibitem{Owens:2004ys}
	\bibinfo{author}{Owens, D.~W.} \& \bibinfo{author}{Lane, E.~B.}
	\newblock \bibinfo{journal}{\bibinfo{title}{Keratin mutations and intestinal
			pathology}}.
	\newblock {\emph{\JournalTitle{J Pathol}}} \textbf{\bibinfo{volume}{204}},
	\bibinfo{pages}{377--85}, \doiprefix\url{10.1002/path.1646}
	(\bibinfo{year}{2004}).
	
	\bibitem{MacFie:2014rm}
	\bibinfo{author}{MacFie, T.~S.} \emph{et~al.}
	\newblock \bibinfo{journal}{\bibinfo{title}{{DUOX2} and {DUOXA2} form the
			predominant enzyme system capable of producing the reactive oxygen species
			{H2O2} in active ulcerative colitis and are modulated by 5-aminosalicylic
			acid}}.
	\newblock {\emph{\JournalTitle{Inflamm Bowel Dis}}}
	\textbf{\bibinfo{volume}{20}}, \bibinfo{pages}{514--24},
	\doiprefix\url{10.1097/01.MIB.0000442012.45038.0e} (\bibinfo{year}{2014}).
	
	\bibitem{Large}
	\bibinfo{author}{Wei, Z.} \emph{et~al.}
	\newblock \bibinfo{journal}{\bibinfo{title}{Large sample size, wide variant
			spectrum, and advanced machine-learning technique boost risk prediction for
			inflammatory bowel disease}}.
	\newblock {\emph{\JournalTitle{The American Journal of Human Genetics}}}
	\textbf{\bibinfo{volume}{92}}, \bibinfo{pages}{1008--1012}
	(\bibinfo{year}{2013}).
	
	\bibitem{amrhein2019scientists}
	\bibinfo{author}{Amrhein, V.}, \bibinfo{author}{Greenland, S.} \&
	\bibinfo{author}{McShane, B.}
	\newblock \bibinfo{title}{Scientists rise up against statistical significance}
	(\bibinfo{year}{2019}).
	
	\bibitem{wasserstein2019moving}
	\bibinfo{author}{Wasserstein, R.~L.}, \bibinfo{author}{Schirm, A.~L.} \&
	\bibinfo{author}{Lazar, N.~A.}
	\newblock \bibinfo{title}{Moving to a world beyond ``p< 0.05''}
	(\bibinfo{year}{2019}).
	
	\bibitem{Maeda:2019db}
	\bibinfo{author}{Maeda, Y.} \emph{et~al.}
	\newblock \bibinfo{journal}{\bibinfo{title}{Fully automated diagnostic system
			with artificial intelligence using endocytoscopy to identify the presence of
			histologic inflammation associated with ulcerative colitis (with video)}}.
	\newblock {\emph{\JournalTitle{Gastrointest Endosc}}}
	\textbf{\bibinfo{volume}{89}}, \bibinfo{pages}{408--415},
	\doiprefix\url{10.1016/j.gie.2018.09.024} (\bibinfo{year}{2019}).
	
\end{thebibliography}

%For data citations of datasets uploaded to e.g. \emph{figshare}, please use the \verb|howpublished| option in the bib entry to specify the platform and the link, as in the \verb|Hao:gidmaps:2014| example in the sample bibliography file.

\section*{Acknowledgements}

This research was partially supported by grants from the Natural Sciences and Engineering Research Council of Canada (NSERC) to H.U. (grant number RGPIN: 2019-05650)  and to L.P.-C.  (grant number RGPIN: 2019-05247). H.M.K. was partially supported by funding from Memorial University’s School of Graduate Studies.

%Acknowledgements should be brief, and should not include thanks to anonymous referees and editors, or effusive comments. Grant or contribution numbers may be acknowledged.

\section*{Author contributions statement}
Conceptualisation H.U. and L.P.-C.; Methodology H.M.K., H.U. and L.P.-C.; 
Analysis  H.M.K. and L.P.-C.; Writing H.M.K., H.U. and L.P.-C.;
 Experiments H.M.K.;
Supervision H.U. and L.P.-C.
%Must include all authors, identified by initials, for example:
%A.A. conceived the experiment(s),  A.A. and B.A. conducted the experiment(s), C.A. and D.A. analysed the results.  All authors reviewed the manuscript. 

\section*{Additional information}

\textbf{Competing interests}
The author(s) declare no competing interests.\\

\textbf{Use of experimental animals, and human participants}
This research did not involve human participants or experimental animals. \\

\textbf{Informed consent}
Not applicable.\\

\textbf{Ethics approval}
Not applicable.\\

%To include, in this order: \textbf{Accession codes} (where applicable);  (mandatory statement). 
%
%The corresponding author is responsible for submitting a \href{http://www.nature.com/srep/policies/index.html#competing}{competing interests statement} on behalf of all authors of the paper. This statement must be included in the submitted article file.

\end{document}